\documentclass[%
twocolumn,
superscriptaddress,
showpacs,
amsmath,amssymb,
longbibliography
]{revtex4-1}
 
\usepackage{graphicx}
\usepackage{dcolumn}
\usepackage{bm}
\usepackage{color}
\usepackage{multirow}
\usepackage[normalem]{ulem}

\def\ket#1{\mathinner{|{#1}\rangle}}
\def\braket#1{\mathinner{\langle{#1}\rangle}}
\begin{document}

\title{%
Unconventional dual 1D-2D quantum spin liquid   \\
revealed by \textit{ab initio}  studies on organic solids family}


\author{Kota Ido}
\author{Kazuyoshi Yoshimi}
\affiliation{Institute for Solid State Physics, University of Tokyo, Kashiwa, Japan}
\author{Takahiro Misawa}
\affiliation{Beijing Academy of Quantum Information Sciences, Haidian District, Beijing 100193, China}
\author{Masatoshi Imada$^*$}
\affiliation{Toyota Physical and Chemical Research Institute, 41-1 Yokomichi, Nagakute, Aichi, 480-1192, Japan}
\affiliation{Waseda Research Institute for Science and Engineering, Waseda University, 3-4-1 Okubo, Shinjuku-ku, Tokyo, 169-8555, Japan} 
\email{imada@g.ecc.u-tokyo.ac.jp}

\begin{abstract}
 Organic solids host various electronic phases. Especially, a milestone compound of organic solid, $\beta'$-$X$[Pd(dmit)$_2$]$_2$ with $X$=EtMe$_3$Sb shows quantum spin-liquid (QSL) properties suggesting a novel state of matter.
However, nature of the QSL has been largely unknown. 
Here, we computationally study five compounds comprehensively with different $X$ using 2D {\it ab initio} Hamiltonians 
and correctly reproduce experimental phase diagram with antiferromagnetic order for $X$=Me$_4$P, Me$_4$As, Me$_4$Sb, Et$_2$Me$_2$As and a QSL
for  $X$=EtMe$_3$Sb without adjustable parameters. 
We find that the QSL for $X$=EtMe$_3$Sb 
exhibits 1D nature characterized by algebraic decay of spin correlation along one direction, while exponential decay in the other direction, indicating dimensional reduction from 2D to 1D.  
The 1D nature indeed accounts for the experimental specific heat, thermal conductivity and magnetic susceptibility. The identified QSL, however, preserves 2D nature as well consistently with spin fractionalization into spinon with Dirac-like gapless excitations
and reveals duality bridging the 1D and 2D QSLs.
\end{abstract}

\maketitle

\noindent
{\bf \large Introduction}

Organic solids offer plenty of playgrounds of 
conspicuous phenomena including superconductivity competing with charge and spin orders, and semi-metallic excitations with Dirac dispersions~\cite{RevModPhys.89.025003,kato2014_BCSJ}. However, the severe competitions of the diverse phases and mechanisms of the phenomena found in complex crystal structures of the organic solids remain challenges of condensed matter physics. 

Despite a large number of atoms in the unit cell of such organic compounds, however, the band structure near the Fermi level is mostly simple consisting of only LUMO (lowest unoccupied molecular orbital) and HOMO (highest occupied molecular orbital) around the Fermi level. Large inter-molecular distance leads to narrow bandwidths while poor screening of Coulomb interaction due to the sparse bands near the Fermi level leads to strongly correlated electron systems with various types of Mott insulators. 

Among them, the QSL phase was
proposed as novel states of matter in two families of correlation driven Mott insulating compounds, namely, $\kappa$-(ET)$_2X$ (ET= bis (ethylenedithio) tetrathiafulvalene) with the anion $X$=Cu$_2$(CN)$_3$\cite{PhysRevLett.91.107001} and $\beta'$-$X$[Pd(dmit)$_2$]$_2$ (dmit=1,3-dithiole-2-thione-4,5-dithiolate) 
with the cation $X$=EtMe$_3$Sb~\cite{Itou2008}, where 
apparent spontaneous symmetry breaking including the antiferromagnetic (AF) order is unusually absent even at several orders of magnitude lower temperatures than the spin exchange interaction. Although the geometrical frustration arising from the triangular lattice structure of the dimerized Pd(dmit)$_2$ or ET molecule hosting a spin-1/2 electronic spin looks important for the absence of the AF order, its {nature and mechanism of the emergence endorsed by a} quantitative estimate are missing. 
In addition to the QSL, if one replaces $X$ with other ions, it shows a variety of phases including the AF order, charge order and valence bond solid~\cite{RevModPhys.89.025003,kanoda2011,kato2014_BCSJ}.  
Because of severe competitions of these phases, {\it ab initio} approaches without adjustable parameters are desired 
particularly for the enigmatic QSL to reach realistic understanding. 

We focus on a typical candidate of QSL, $X$[Pd(dmit)$_2$]$_2$. In fact, it shows~\cite{Itou2008,Itou2010} smaller broadening of NMR spctra than $\kappa$-(ET)$_2$Cu$_2$(CN)$_3$\cite{RevModPhys.89.025003,PhysRevB.73.140407} implying smaller effects of extrinsic spatial inhomogeneity. 
At low temperatures of the QSL material, EtMe$_3$Sb[Pd(dmit)$_2$]$_2$, the specific heat~\cite{Yamashita2011} and the thermal conductivity $\kappa$~\cite{Yamashita2010} are reported to be proportional to temperature $T$, although controversies 
exist for the latter~\cite{Yamashita2020,PhysRevX.9.041051,PhysRevLett.123.247204}. The magnetic susceptibility stays at a nonzero constant~\cite{Watanabe2012}. These are roughly similar to the conventional Fermi-liquid metal but of course the electrons are frozen in the Mott insulator 
and the spin degrees of freedom must be responsible for them.
In addition, the relaxation rate of the nuclear magnetic resonance (NMR), $1/T_1$ seems to be scaled by $T^2$ in the range 0.05K$\le T<1$K~\cite{Itou2010} in contrast to the conventional Fermi liquid behavior $\propto T$.  It is then desired to gain insights into the experimental consistency from {\it ab initio} calculations. 

Here, we thoroughly study purely {\it ab initio} electronic Hamiltonians without any adjustable parameters for five dmit compounds  using the experimental structure without optimizing lattice structures except
for positions of hydrogen atoms. This {\it ab initio} study allows us to correctly reproduce the overall experimental phase diagram at low temperatures of 
$\beta'$-$X$[Pd(dmit)$_2$]$_2$ with $X$=Me$_4$P, Me$_4$As, Me$_4$Sb, Et$_2$Me$_2$As exhibiting the AF order and with $X$=EtMe$_3$Sb exhibiting QSL phases. 
We do not consider $\beta'$-Et$_2$Me$_2$Sb[Pd(dmit)$_2$]$_2$, because the observed charge order with the simultaneous spontaneous lattice distortion in this compound~\cite{nakao2005structural,tamura2005spectroscopic} requires structural optimization, which is out of the scope of this paper.
Then we conclude that the QSL phase has a character of primarily 1D spin liquid but combined with 2D nature, which accounts for the experimental properties consistently.
{From the study, we attempt to pin down the nature of QSL observed in the experiments, which is one of the central issues in condensed matter physics.} \\

\begin{figure}[]
  \begin{center}
    \includegraphics[width=8cm,clip]{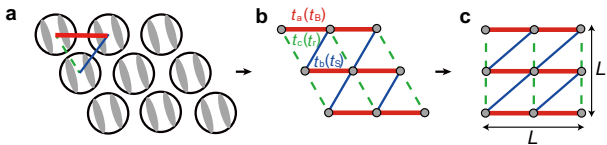}
  \end{center}
  \caption{
    {\bf Lattice structure of $\beta'$-$X$[Pd(dmit)$_2$]$_2$.}
{\bf a}: Schematic triangular structure of $\beta'$-$X$[Pd(dmit)$_2$]$_2$ consisting of dimerized Pd(dmit)$_2$ molecules, where a Pd(dmit)$_2$ molecule is depicted as a long gray oval. 
{\bf b}: Modeled triangular lattice with three different electronic transfers. The strongest, middle and weakest transfers, $t_a$ (on the red bond $\tilde{a}$), $t_b$ (on the blue bond $\tilde{b}$), and $t_c$ (on the green bond $\tilde{c}$) (they correspond to $t_{\rm B}$, $t_{\rm S}$ and $t_{\rm r}$, respectively in the literature\cite{kato2012cation}). 
{\bf c}: Deformed structure on a $L \times L$ lattice with the system size $N_s=L^2$ used in the present calculation. 
}
\label{fig1}
\end{figure}
\noindent

\noindent
{\bf \large Results} \\
{\bf  \textit{Ab initio} framework. } 
Before going into our results, we summarize our framework of the calculation for the sake of  readers to understand definitions of quantities, which we will analyze. 
We use {\it ab initio} low-energy effective Hamiltonian consisting of the half-filled HOMO band crossing the Fermi level 
derived for $\beta'$-$X$[Pd(dmit)$_2$]$_2$~\cite{Nakamura2012,PhysRevResearch.2.032072,Misawa2021}. 
For the procedure and the obtained parameters for the Hamiltonians, see Methods.
It is built on the weakly coupled two-dimensional layers and the Pd(dmit)$_2$ dimer forms an anisotropic triangular lattice within a layer. We assign the $\tilde{a}$, $\tilde{b}$ and $\tilde{c}$ bonds, for which the Hamiltonian presented below (Eq.~(\ref{Hamiltonian})) contains the transfer parameters $t_a$, $t_b$ and $t_c$, respectively, as is illustrated in the middle panel of Fig.~\ref{fig1}. The bonds are chosen so that $t_a$, $t_b$ and $t_c$ are ordered from the stronger to weaker amplitudes. In the actual calculation, we take a deformed lattice structure illustrated in the right panel of Fig.~\ref{fig1} just for computational simplicity. 
We sometimes use $x$ and $y$ axes for the plot of the 2D Brillouin zone and note that $x$ and $y$ correspond to $\tilde{a}$ and $\tilde{c}$ directions, respectively. (Do not confuse the crystallographic $a, b$ and $c$ axes with the present assignment of $\tilde{a}$, $\tilde{b}$ and $\tilde{c}$ axes.)

\begin{figure*}[]
  \begin{center}
    \includegraphics[width=12cm,clip]{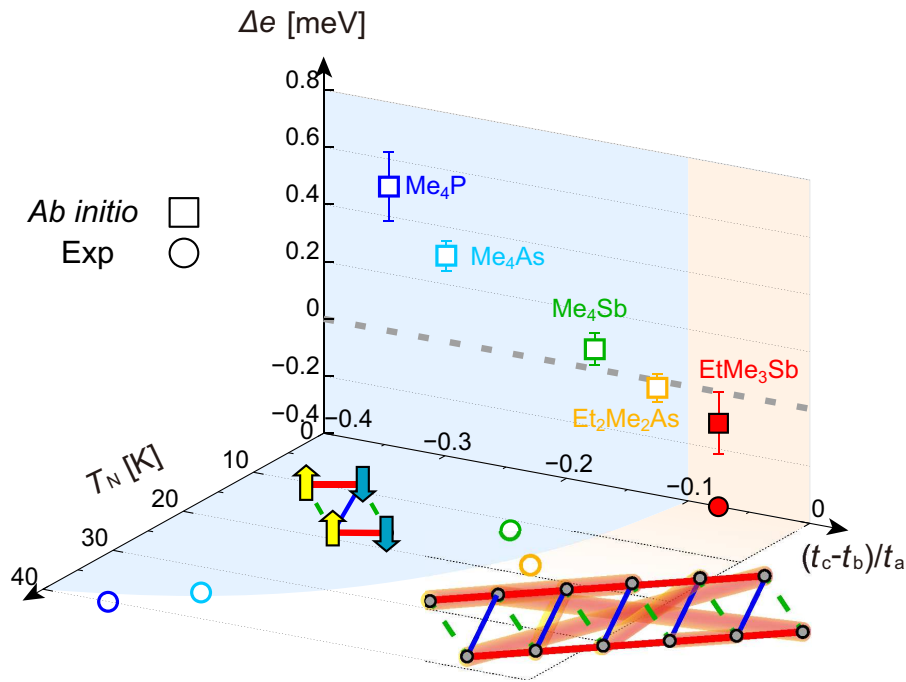}
  \end{center}
  \caption{
    {\bf Ground-state phase diagram revealed by $ab$ $initio$ simulations.}
Stability of electronic phases for {\it ab initio} Hamiltonians of dmit compounds, plotted for $X$=Me$_4$P, Me$_4$As, Me$_4$Sb, Et$_2$Me$_2$As, and EtMe$_3$Sb aligned from the left (small $(t_c-t_b)/t_a$) to the right. 
{Vertical plane: Square symbols show the energy difference $\Delta e=e_{\rm QSL}-e_{\rm AF}$ per site between the energy of the QSL and that of the AF phases calculated for $L=16$, where the energies of two lowest states, the QSL and the collinear AF states are compared. 
Negative} $\Delta e$ indicates that the QSL is the ground state. 
The results with error bars was obtained after the energy variance extrapolation described in Methods.
{Bottom plane: Circles show the material dependence of the N\'{e}el temperature $T_N$ observed in the experiments~\cite{Nakamura2001,Fujiyama2019}.
 For Et$_2$Me$_2$As and Me$_4$Sb, coexistence of QSL and AF is suggested.
For the both planes, open symbols show the AF states and the filled ones are the QSL. Blue and orange shades represent the region of the collinear AF and the QSL phases, respectively for the parameter $(t_c-t_b)/t_a$. }
}
\label{gspd}
\end{figure*}

 Because of very weak interlayer hopping (the largest interlayer hopping is less than 0.8 meV), the 2D plane is sufficient to capture physics of our interest about the ground state here. Then our Hamiltonian is built on the two-dimensional plane of the dmit salts.
We take into account the effect of interlayer interaction through the dimensional downfolding method established in organic solids and iron-based superconductors~\cite{Nakamura2012} (see Methods).  
The maximally localized Wannier function of the HOMO orbital, whose band crosses the Fermi level is constructed for the molecular orbital\cite{marzari1997,souza2001}. The lattice structure in the simulation is depicted later in Fig.~\ref{fig6},
which is, despite the  deformation of the shape, able to sufficiently describe the {\it ab initio} Hamiltonian as the network of the Pd(dmit)$_2$ dimers, where a Wannier orbital is extended over a dimer.  

The resultant {\it ab initio} single-band effective Hamiltonian derived in Ref.~\cite{PhysRevResearch.2.032072,Misawa2021} has the form 
\begin{eqnarray}
  {\cal H}  =  -  \sum_{\sigma} \sum_{i,j} t_{ij} c_{i \sigma}^{\dagger}c_{j \sigma} +  \sum_i U n_{i \uparrow} n_{i \downarrow} \nonumber \\
+  \sum_{i<j} V_{ij} (n_i-1)(n_j-1), 
\label{Hamiltonian}
\end{eqnarray}
where $i$, $j$ represent the dimer indices, and $c_{i \sigma}^{\dagger}$ ($c_{i \sigma}$) is the creation (annihilation) operator of electrons with spin $\sigma$ (=$\uparrow$ or $\downarrow$) at the $i$-th Wannier orbital, and the number operator is $n_i = \sum_{\sigma} n_{i\sigma}$ with $n_{i\sigma}=c^{\dagger}_{i\sigma} c_{i\sigma}$. Here, $t_{ij}$ is the hopping parameters depending on the relative coordinate vector $\bm{r}_i -\bm{r}_j$, where $\bm{r}_i$ is the position vector of the center of the $i$-th Wannier orbital. In the present study, for $t_{ij}$ and $V_{ij}$, we retain the nearest neighbor pair of $\bm{r}_i$ and $\bm{r}_j$ in each direction of $\tilde{a}$, $\tilde{b}$ and $\tilde{c}$ as is illustrated in Fig.~\ref{fig1}. Here, $U$ and $V_{ij}$ are the screened on-site and off-site Coulomb interactions, respectively, as is illustrated in Fig.~\ref{fig6}. Note that the Hamiltonian parameters of both transfer and interaction terms  contain neither adjustable parameters nor fitting and are determined solely by using the experimental lattice structure at low temperatures and by following the established procedure of the maximally localized Wannier functions. The spatial ranges of the transfer and the interaction, which we take are sufficient to reach the convergence for the present study and the values at longer distance are small. We also ignore the direct exchange interactions because they are at most 3 meV and we expect it does not alter our conclusions. See Methods 
for the parameter values of Hamiltonian (\ref{Hamiltonian})  used in the present study.
 
We apply 
a variational Monte Carlo (VMC) method~{\cite{Tahara2008,MISAWA2019447,Nomura2017}} to the {\it ab initio} Hamiltonians to reach  highly accurate ground states.
For details of the numerical method, see Methods. \\

\noindent
{\bf  Phase Diagram. }
Figure \ref{gspd} is one of our central results of this paper, showing the agreement between the experimental (bottom plane) and the calculated (vertical plane)  material dependences of the low-temperature phases.
In the calculated results, the collinear AF and the QSL states are the two lowest energy states among severely competing various candidates all through the five compounds studied. 
Therefore, we plot the calculated energy difference between these two in the vertical plane. 
 We find that the experimental QSL is reproduced in the calculated ground state for $X$=EtMe$_3$Sb. 
Aside from a severe competition for $X$=Et$_2$Me$_2$As, which is indeed suggested by the experimental coexistence of AF and QSL.  
Fig.~\ref{gspd} shows that our {\it ab initio} results successfully reproduce the AF ground state observed in the experiments of $X$=Me$_4$P, Me$_4$As, and Me$_4$Sb, as shown in the bottom plane. The AF state has the Bragg peak at $(\pi,0)$ in the notation of Fig.\ref{fig1}. 
The AF stability was recently studied by the quasi-1D approach based on the random phase approximation~\cite{PhysRevMaterials.5.084412}.

Note that the abscissa of Fig.~\ref{gspd} shows only $(t_c-t_b)/t_a$ dependence among the {\it ab initio} parameters, while the plots of the five materials are the results of computation by using the full {\it ab initio} parameters. The overall monotonic dependence of $\Delta e$ shows that $(t_c-t_b)/t_a$ is indeed a principally important parameter for evaluation of the phase diagram. 
The experimental N\'{e}el temperature in the bottom plane is also consistently ordered with $(t_c-t_b)/t_a$ and further supports the relevance and accuracy of the {\it ab initio} effective Hamiltonian, because it is natural to expect that the N\'{e}el temperature is linked to the relative stability of the ground state. Within the parameter control of $(t_c-t_b)/t_a$, the transition between the AF and QSL phases is a clear first-order transition, represented by the level crossing.
The importance of the full anisotropy of the transfer in the inequilateral triangular lattice to stabilize the QSL phase was pointed out by the projected BCS study of the Hubbard-type model by using the {\it ab initio} hopping~\cite{PhysRevB.88.155139}.\\  

\begin{figure}[htp]
  \begin{center}
    \includegraphics[width=8cm,clip]{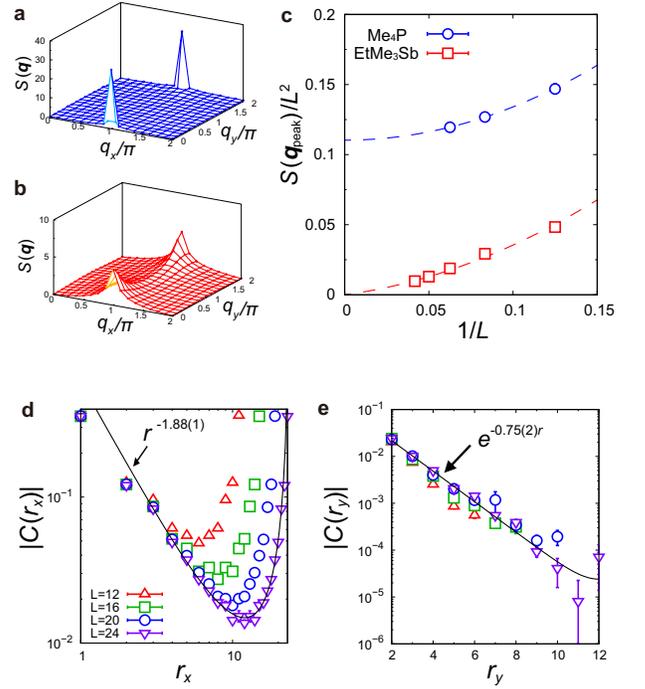}
  \end{center}
  \caption{
    {\bf Spin-spin correlation function $C(\bm{r})$ and spin structure factor $S(\bm{q})$ (Fourier transform of $C(\bm{r})$) obtained by using the VMC method.}  
    {\bf a}: $S(\bm{q})$ of the AF state for $X$=Me$_4$P with $L=16$ on a $L\times L$ lattice.
    {\bf b}: $S(\bm{q})$ of the QSL for $X$=EtMe$_3$Sb with $L=16$.
    In {\bf a} and {\bf b}, the unit of the momentum $q_x$ and $q_y$ is $\pi$.
    {\bf c}: Size dependences of the peak value of $S(\bm{q})$ denoted as $S(\bm{q}_{\rm peak})$ in the AF state for $X$=Me$_4$P and the QSL state for $X$=EtMe$_3$Sb. 
    Dashed lines are guides for the eye.
    The AF state scales to a nonzero $S(\bm{q}_{\rm peak})/N_s$ in the thermodynamic limit, evidencing for the AF long-range order.
On the other hand, 
extrapolation to zero in the thermodynamic limit for the QSL state confirms the non-magnetic state.
    {\bf d}: Log-log plot of $C(\bm{r}=(r,0))$ for $X$=EtMe$_3$Sb. Black thin line is the fitting curve 
$C_{\rm fit}(\bm{r}) \propto \sum_{n}|nL+r|^{-p}$ with $n$ being integer by taking account of the antiperiodic boundary condition,
which is fit by $p=1.88$ for $L=24$.
    We use data points for $2 < r < L-2$ when we obtain the fitting curve.
    {\bf e}: Semi-log plot of $C(\bm{r}=(0,r))$ for $X$=EtMe$_3$Sb. Black thin line is the fitting curve 
$C_{\rm fit}(\bm{r}) \propto \sum_{n}[\exp(-|nL+r|/\xi_\perp)$
with $\xi_\perp=1.33$ for $L=24$. Error bars are obtained from the Monte Carlo sampling and the variance extrapolation, details of which are described in Methods. 
The length unit is the lattice constant in panels {\bf d} and {\bf e}.
In panels {\bf c}-{\bf e}, error bars obtained in the VMC calculation are plotted and if not they are smaller than the symbol size. 
}
\label{fig2}
\end{figure}

\noindent
{ \bf Nature of quantum spin liquid and antiferromagnetic state. } 
Figure~\ref{fig2} shows the spin structure factor $S(\bm{q})$ (the spin correlation in the Fourier space) for $X$=Me$_4$P 
as a representative case of the AF ordered compounds and for the QSL compound, $X$=EtMe$_3$Sb (see Methods for details of the calculation method and definition of physical quantities).
For the AF state of $X$=Me$_4$P, a strong sharp peak at $\bm{q}=(\pi, 0)$ 
indeed indicates the conventional collinear {2D} AF order. The {2D} ordering pattern is illustrated in Fig.~\ref{gspd}. Other compounds,  $X$=Me$_4$As, Me$_4$Sb, and Et$_2$Me$_2$As also show the same type of the order.
In the QSL state for EtMe$_3$Sb, $S(\bm{q})$ also has a peak at the same momentum $\bm{q}=(\pi, 0)$,
but with a substantially reduced height.
Interestingly, a ridge line along $q_x=0$ emerges in $S(\bm{q})$ of the QSL state, implying anisotropy between the $x$ and $y$ directions (namely between the chain ($\tilde{a}$) and interchain ($\tilde{b}$) directions). {Such anisotropy is in sharp contrast with the isotropic order in the AF phase. 
Furthermore}, the feature of the anisotropy is largely different from the previous numerical studies for the anisotropic triangular Hubbard or Heisenberg model~\cite{PhysRevB.89.235107}.
Figure \ref{fig2}(c) shows the size dependence of the peak value of $S(\bm{q})$ divided by the system size $N_s=L^2$. 
Because the AF long-range order is signaled by a nonzero $S(\bm{q})/N_s$  in the thermodynamic limit $1/L\rightarrow 0$, AF order exists for $X$=Me$_4$P and it is absent for $X$=EtMe$_3$Sb.

The spin-spin correlations in real space $C(\bm{r})$ for the QSL state of $X$=EtMe$_3$Sb
shown in Figure \ref{fig2}(d) with the log-log plot in the $x$ (namely, $t_a$) direction indicates a power law $C(\bm{r})\propto |\bm{r}|^{-p}$ with $p= 1.88\pm 0.01$ suggesting the algebraic QSL with the gapless excitation. 
On the other hand, Fig. \ref{fig2}(e) shows a semilog plot of the correlations in the $y$ direction, for which the exponential form $C({\bm r})\propto \exp[-|{\bm r}|/\xi_{\perp}]$ with the correlation length $\xi_{\perp}\sim 1.33\pm 0.04 $ is appropriate and thus the excitation is gapped in the interchain ($y$) direction at least within the system size studied here. Of course, the same exponential decay of the correlation is observed in the $t_b$ (namely $\tilde{b}$)  and $t_c$ (namely, $\tilde{c}$) directions.
The anisotropic correlation develops the ridge in the $q_y$ dependence of $S(\bm{q})$.

\begin{figure}[bt!]
  \begin{center}
    \includegraphics[width=8cm,clip]{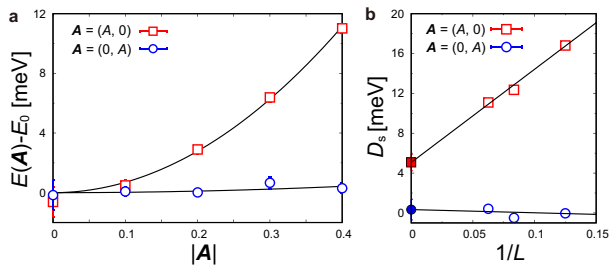}
  \end{center}
  \caption{
    {\bf Spin Drude weight $D_{\rm s}$ for $X$=EtMe$_3$Sb obtained by using the VMC method.} 
    {\bf a}: Vector potential dependence of the total energy with $L=16$ for the QSL of $X$=EtMe$_3$Sb. 
        Squares and circles are the results in $x$ and $y$ directions, respectively.
        $D_{\rm s}$ is obtained by fitting the total energy shown here in the form 
$E(\bm{A}) = 2\pi D_{\rm s} |\bm{A}|^2 + E_0$ as shown as the dashed lines.
The energy unit is eV.
We find a substantial curvature of the energy in $x$ direction for the QSL, 
indicating a large $D_{\rm s}$ for the QSL.   
In all the data plot, error bars obtained in the VMC calculation are plotted and if not they are smaller than the symbol size. 
    {\bf b}: System size dependence of $D_{\rm s}$ in $x$(squares)  and $y$ (circles) directions for the QSL.
    Filled symbols represent $D_{\rm s}$ in the thermodynamic limit, which are obtained after the linear extrapolation.
Error bars arising from the fitting or the extrapolation are plotted and if not they are smaller than the symbol size. }
\label{fig4}
\end{figure}

Finally, the spin Drude weight $D_{\rm s}$ calculated in the QSL state for $X$=EtMe$_3$Sb 
is shown in Fig.~\ref{fig4}.  
The size dependence of $D_{\rm s}$ for the QSL shown in Fig. \ref{fig4}(b)
assures that $D_{\rm s}$ is nonzero and large in the thermodynamic limit, consistently with the power-law decay of the spin correlation in the $x$ direction.
In contrast, the spin Drude weight for the interchain $y$ component
is vanishing. 

All the results support that the QSL with one-dimensional and gapless excitation is realized for $X$=EtMe$_3$Sb, which is our second central result of the paper.\\ 

\noindent
{\large \bf Discussion \\} 
The QSL of $\beta'$-EtMe$_3$Sb[Pd(dmit)$_2$]$_2$ is characterized by the gapless spin excitation and associated algebraic power-law decay of the spin correlation in the direction of the largest transfer $t_a$ (namely, along the chain direction). 
The power $p \sim 1.9$ for $C(\bm{r})\propto \exp[i\bm{Q}\cdot\bm{r}]\bm{r}^{-p}$ is similar to the value obtained for the QSL in the square-lattice $J_1$-$J_2$ Heisenberg model with the nearest ($J_1$) and the next-nearest-neighbor ($J_2$) exchange interactions, where $p\sim 1.4$-$1.7$~\cite{NomuraJ1J2_2021}. However, in contrast to isotropic 2D spin correlation in the $J_1$-$J_2$ model, the correlation decays exponentially in the interchain direction 
with the correlation length $\xi_{\perp}\sim 1.3$ lattice constant.  
The anisotropy makes the peak height of $S({\bm q})$ non-divergent in contrast with the 2D $J_1$-$J_2$ model.

One might regard the present QSL as essentially the same as the 1D spin liquid~\cite{PhysRevB.89.235107,PhysRevB.74.012407,PhysRevB.74.014408} smoothly connected to an effective 1D Heisenberg or Hubbard model. 
However, the revealed properties are not so simple.
First, the nonzero correlation length in the interchain direction makes a prominent peak at $(\pi,0)$ in $S(\bm{q})$ commonly to the case of the 2D $J_1$-$J_2$ Heisenberg model. 
In the decoupled arrays of 1D Heisenberg chains, one would expect an equal-height ridge along the line $(\pi,0)$-$(\pi,\pi)$ with logarithmically divergent height where $p=1$ and without $q_y$ dependence. In the present QSL, the ridge exists but the height is not divergent even at $(\pi,0)$ because $p>1$. For the moment, we leave the issue about the possible presence or absence of the transition between the pure isolated chain and the present case for the future study.
Substantial reduction of $S(\bm{q})$ toward $(\pi,\pi)$ from $(\pi,0)$, namely partially 2D-like correlation implies a small but nonzero dispersion of the excitation in the interchain direction, which may make an energetic hierarchy.

\begin{figure}[h!]
  \begin{center}
    \includegraphics[width=8cm,clip]{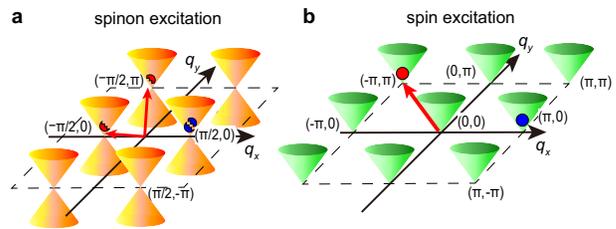}
  \end{center}
  \caption{
{\bf Schematic illustration of the gapless excitation.}  
{\bf a}: Gapless structure of the spinon in Brillouin zone inferred from the structure of the variational wavefunction discussed in Methods. 
{\bf b}: Visible triplet excitation constructed from two spinon excitations.  Two examples of gapless excitations are shown: Two red spinons, one at  $(-\pi/2,\pi)$ and the other at $(-\pi/2,0)$  in {\bf a} can make visible gapless triplet excitation shown in {\bf b} at $(-\pi,\pi)$. Two blue spinons at $(\pi/2,0)$ in {\bf a} can make visible gapless triplet excitation shown in {\bf b} at $(\pi,0)$.
}
\label{fractionalization}
\end{figure}

\begin{figure}[bt!]
  \begin{center}
    \includegraphics[width=8cm,clip]{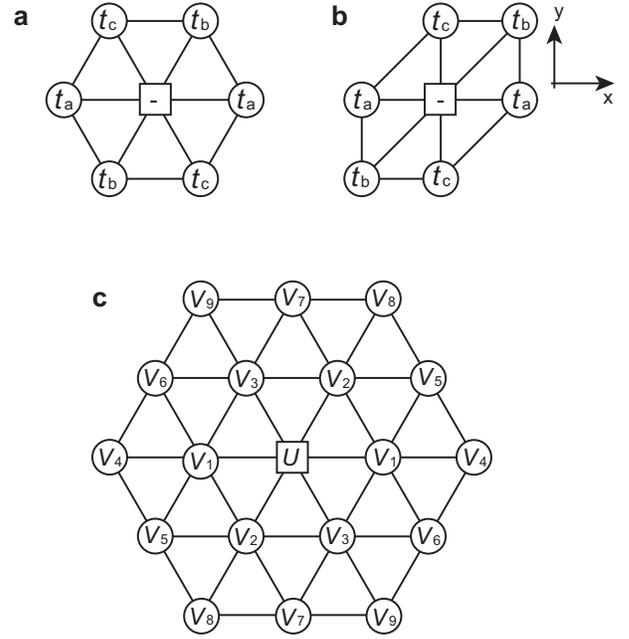}
  \end{center}
  \caption{
    {\bf Definitions of {\it ab initio} parameters and the form of the lattice for $X$[Pd(dmit)$_2$]$_2$.}
    {\bf a}: Definitions of hopping amplitudes. $t_a > t_b > t_c$ is satisfied in all the single band effective Hamiltonians we studied.
    The square denotes the reference site and $t_n$ ($n=a$,$b$, and $c$) on circles denotes the hopping amplitudes from the reference site.
    The values are listed in Table \ref{tab:abval}.
    {\bf b}: Definitions of deformed triangular lattice and $x-y$ axes.
    {\bf c}: Definitions of screened Coulomb interactions on the lattice corresponding to non-deformed lattice in {\bf a}. $U$ and $V_n$ (n=1-9) on circles denote the screened Coulomb interactions from the reference site (square in the center). In the actual calculation, the same deformed structure as {\bf b} is used.
    The values are listed in Table \ref{tab:abval}.
}
\label{fig6}
\end{figure}

For the 2D $J_1$-$J_2$ model, it was suggested that the gapless excitation is well accounted for by the fractionalization of a spin into two spin-$1/2$ spinons, where the spinon has the Dirac-like linear dispersion at $(\pm\pi/2, \pm\pi/2)$~\cite{Ferrari_2018,NomuraJ1J2_2021}.  
Let us discuss the possibility of the fractionalization for the present QSL of $\beta'$-EtMe$_3$Sb[Pd(dmit)$_2$]$_2$. 
To gain insight into the possible existence of spinon and to elucidate its dispersion in the present QSL, we have analyzed the structure of our wavefunction for the QSL ground state of $\beta'$-EtMe$_3$Sb[Pd(dmit)$_2$]$_2$ as is described in Methods. 
The result shows that the elementary excitation is consistently described by the spinon born out as a consequence of the fractionalization of a spin, where the spinon has Dirac-type gapless dispersion around  $(\pm\pi/2,0)$ and $(\pm\pi/2,\pi)$ as is schematically illustrated in Fig.~\ref{fractionalization}. Because the measurable spin is constructed from the two-spinon excitation, the gapless points for the triplet excitations appear at around $(0,0)$, $(\pm\pi,0)$, $(0,\pm\pi)$, and $(\pm\pi,\pm\pi)$. 
This structure has an essential similarity to the 2D $J_1$-$J_2$ Heisenberg model~\cite{NomuraJ1J2_2021}, implying the 2D nature of the QSL.

However, the exponential decay of the spin correlation and vanishing spin Drude weight in the interchain direction suggest a very anisotropic and small or incoherent dispersion away from the nodal point in the interchain direction in contrast to the isotropic conic dispersion for the 2D $J_1$-$J_2$ model, generating the QSL of roughly 1D-like nature.

By considering the one dimensional anisotropy,
it is insightful to compare the present QSL with that of the 1D antiferromagnetic Heisenberg model defined by
\begin{eqnarray}
H=J\sum_i [{S}_i^x{S}_{i+1}^x+{S}_i^y{S}_{i+1}^y+{S}_i^z{S}_{i+1}^z]
\label{1DHeisenberg}
\end{eqnarray}
with $J>0$ for the spin half operator $\bm{S}_i=({S}_i^x, {S}_i^y, {S}_i^z) $ at site $i$.
The specific heat of the 1D Heisenberg model is given by the $T$-linear coefficient $\gamma=2k_{\rm B}^2 N_{\rm A}/3J$ above 1K~\cite{Takahashi1973}, where $k_{\rm B}$ is Boltzmann constant and $N_{\rm A}$ is Avogadro constant. 
If we employ the strong coupling expansion for the {\it ab initio} Hamiltonian, the superexchange interaction $J_a$ for the two neighboring spins in the $\tilde{a}$ (namely, $t_a$) direction is estimated as $J_a=4t_a^2/(U-V_1)\sim 0.031$ eV for $\beta'$-EtMe$_3$Sb[Pd(dmit)$_2$]$_2$, 
by using the {\it ab initio} parameters listed in Table \ref{tab:abval} of Methods. 
Then $\gamma$  is estimated as
$\gamma\sim 15$ mJ K$^{-2}$ mole$^{-1}$, comparable with the experimental value for $\beta'$-EtMe$_3$Sb[Pd(dmit)$_2$]$_2$, $\gamma^{\rm exp}\sim 15-20$ mJ K$^{-2}$mole$^{-1}$~\cite{Yamashita2011,PhysRevLett.123.247204}.

The uniform magnetic susceptibility of the 1D Heisenberg model is $\chi_{\rm 1D}=g^2\mu_{\rm B}^2/(J_a\pi^2)$ with the $g$-factor and the Bohr magneton $\mu_{\rm B}$~\cite{Griffiths1964,Yang1966}.
By using $J_a\sim 0.031$eV,  and $g\sim2$, we obtain $\chi/\mu_{\rm B}^2\sim 13$ eV$^{-1}$, while the experimental value is $\chi/\mu_{\rm B}^2 \sim 12-24$ eV$^{-1}$ per the dimer~\cite{Itou2008,Watanabe2012}, 
which is consistent with each other by considering the experimental uncertainty.

Frequency dependent thermal conductivity $\kappa$ of the 1D Heisenberg model is estimated to be  
$\kappa^{\rm 1D}(\omega)=\kappa_{\rm s}\delta(\omega)$
with 
$\kappa_{\rm s}/{T}={\pi^3J}/(6a\hbar^2)$,
where $a$ is the lattice constant\cite{Klumper2002}.
If the delta function is replaced by the Lorentzian $\omega_{\rm W}/\pi(\omega^2+\omega_{\rm W}^2)$ to take into account the lifetime $\tau=2\pi/\omega_{\rm W}$, 
we obtain 
$\kappa_{\rm s}^{\rm 1D}(\omega=0)/T=\pi^2 J_a/(6a\hbar^2 \omega_{\rm W})$.
By considering $a\sim 10^{-9}$m and {$J_a\sim 0.031$}eV, 
$\kappa_{\rm s}^{\rm 1D}(\omega=0)/T$ is estimated as 
{$7.7\times 10^{22}/\hbar \omega_{\rm W}$}[Jms]$^{-1}$.
Then if the experimental value reported in Ref.~\cite{Yamashita2010} is employed, {and the one-to-one correspondence between the 1D Heisenberg and the experimental value is assumed}, $\kappa_{\rm exp}^{\rm dmit}/T\sim0.2/k_{\rm B}^2$[Jms]$^{-1}$ corresponds to 
the carrier relaxation time 
{$\tau=2\pi/ \omega_W \sim 9.0\times 10^{-12}$}sec.
A simple expectation of the spin velocity $v_{\rm s}=J_aa/\hbar$ results in the mean free path $\ell_{\rm s}=v_{\rm s} \tau=0.42\mu$m, which is a value {comparable to} the reported estimate $\sim 1\mu$m~\cite{Yamashita2010}. 
{More precisely, the experimental value interpreted by the 1D Heisenberg model suggests $\hbar \omega_{\rm W}/J\sim 0.015$ implying small damping  in the order of $10^{-2}$ for the propagation induced plausibly by the spinon dynamics driven by the energy scale of $J$}. 
Therefore, although controversies exist as is mentioned above, if a large thermal conductivity experimentally reported is intrinsic, it can be essentially accounted for by the interpretation of 1D-like QSL.
The obtained $\ell_{\rm s}$ is {substantially} longer than that of the inorganic compound Cs$_2$CuCl$_4$\cite{PhysRevResearch.1.032022}, where a 1D QSL was claimed. The present result implies that the 1D QSL found here has potentially much longer $\tau$. 
Then all of the above thermodynamic and transport properties can be interpreted by the essentially 1D-like QSL found here.

The spin-lattice relaxation time $T_1$ 
{reported as} scaled by $1/T_1\propto T^2$ below 1K~\cite{Itou2010}, implying the point-like gapless triplet excitation 
looks contradicting the 
$T$-linear 
specific heat and nonzero magnetic susceptibility {compatible with the constant density of states}. 
However, a consistent picture {may emerge,} if 
a spinon with a highly anisotropic nodal and gapless dispersion exists at $\sim (\pm \pi/2,0)$ and $\sim (\pm \pi/2,\pi)$,
where the 2D nature could be detectable only at low temperatures below 1K and by measuring the gapless momenta with an appropriate form factor. 
{The power of the dispersion around the node in the chain direction is an intriguing issue but is beyond the scope of the present paper.}

It turned out that the Dirac-type gapless excitation for the 2D QSL can be connected to the vanishing spinon dispersion along the Fermi line for $T>1$ K, which behaves as if it is essentially the 1D QSL. 
$\beta'$-EtMe$_3$Sb[Pd(dmit)$_2$]$_2$ is located in this parameter space close to the 1D QSL.
Furthermore, the controversy and sensitivity of the thermal conductivity~\cite{Yamashita2020,PhysRevX.9.041051,PhysRevLett.123.247204} are well understood from the one-dimensional nature:
The exponential decay of the spin correlation and the vanishing Drude weight imply the vanishing thermal conductivity in the interchain direction.  Then even small angle misalignment of the single crystal either in samples or in measurements causes serious reduction of $\kappa$ and serious sample or measurement dependence. Effects of randomness may also seriously disturb the ideal behavior of the 1D-like spin liquid~\cite{PhysRevX.9.041051}.
Observed very sensitive dependence 
might be able to be interpreted from the 1D nature of the QSL, which has also been pointed out in Ref.~\cite{PhysRevMaterials.4.044403}. (See also Methods~\ref{app:strong_coupling})

In summary, we have studied the family of dmit organic salts by using the {\it ab initio} Hamiltonians of 5 compounds. 
The obtained low-temperature phases are consistent with the experimental reports as the AF state for $X$=Me$_4$P, Me$_4$As, Me$_4$Sb, Et$_2$Me$_2$As and the QSL for EtMe$_3$Sb. The relative stabilities of the four AF compounds increasing with decreasing $t_c-t_b$ are correctly reproduced in the {\it ab initio} calculation in agreement with the experimental trend.
{Thanks to this firm correspondence,} the nature of the QSL {in the real material} is identified as the 1D-like spin liquid: The spin correlation decays exponentially and the spin Drude weight vanishes in the interchain direction. Though a controversy exists in the experimental reports of the thermal conductivity, the specific heat, the susceptibility and potentially the thermal conductivity show overall consistency with the experimentally observed values essentially in terms of the 1D QSL. However, the signature of the 2D properties is found in a prominent peak of the spin structure factor at $(\pi,0)$ and in the structure of the ground-state wave function itself.  In addition, the exponent of the algebraic correlation is different from the known 1D QSL found in the Heisenberg or Hubbard model. 
The temperature dependence of NMR $T_1^{-1}$ at $T<1$ K ($\propto T^2$) 
{supports vanishing density of states of spin excitation around zero energy and implies the existence of highly anisotropic but point like nodes consistently with the present results.} 
The present QSL {offers} a unified picture that bridges the 1D and 2D QSL.

It is desired to calculate the dynamical spin structure factor in the future to more directly clarify the spin dynamics with nodal excitation suggested by the structure of the wave function.
An intriguing issue is the relation to recent studies on the charge dynamics above 2K for $\beta'$-EtMe$_3$Sb[Pd(dmit)$_2$]$_2$~\cite{PhysRevB.97.035131}. It might be associated with essentially the 1D-like spin liquid, where the dynamical singlet formation may couple to the dimerization fluctuation of the lattice. The spin transport can be measured by attaching ferromagnetic metal leads and by estimating the difference of the spin conduction between the cases of parallel and anti-parallel magnetization for the anode and cathode. The ac response may also help to estimate the spin Drude weight studied in the present study
without the sensitivity to the misalignment or randomness.  These are left for future studies to stringently verify the relevance of the present finding in the collaboration of experiment and theory.\\

\vspace{5mm}
\noindent
{\bf \large Methods \\}
\label{app:me} 
\begin{table}[h]

\caption{
{\bf Derived parameters of the {\it ab initio} effective Hamiltonian for $X$[Pd(dmit)$_2$]$_2$.} 
$t_n$ and $V_n$ represent the hopping parameters and Coulomb interactions, respectively, defined in Fig.~\ref{fig6}. The units of $t_n$, $U$ and $V_n$ are meV. After the dimensional downfolding, all the interaction parameters $U$ and $V_n$ should be reduced with the amount of 180 meV.  
}
\begin{center}
\scalebox{1.25}{
\footnotesize
\begin{tabular}{l|ccccc}
\hline \hline
   &  \multicolumn{5}{c}{Cation $X$} \\
   &  Me$_4$P & Me$_4$As & Me$_4$Sb & Et$_2$Me$_2$As & EtMe$_3$Sb \\ 
\hline 
   $t_a$ & 60.6 & 63.0 & 56.1 & 55.6 & 57.1\\
   $t_b$ & 51.8 & 49.5 & 45.7 & 43.8 & 44.6\\
   $t_c$ & 30.8 & 30.7 & 35.8 & 36.8 & 40.3\\
   $U$   & 793 & 798 & 824 & 851 & 840 \\
   $V_1$ & 414 & 414 & 411 & 431 & 413\\
   $V_2$ & 427 & 428 & 428 & 449 & 434\\
   $V_3$ & 379 & 381 & 385 & 406 & 390\\
   $V_4$ & 295 & 293 & 288 & 307 & 289\\
   $V_5$ & 323 & 322 & 317 & 337 & 320\\
   $V_6$ & 295 & 294 & 291 & 310 & 293\\
   $V_7$ & 307 & 308 & 307 & 326 & 311\\
   $V_8$ & 314 & 314 & 310 & 330 & 315\\
   $V_9$ & 278 & 278 & 276 & 295 & 279\\
\hline \hline
\end{tabular}
}
\end{center}
\label{tab:abval}
\end{table}

\noindent
{\bf Effective Hamiltonian parameters. }
\label{app:ham} 
We have used the parameters of the effective Hamiltonian (\ref{Hamiltonian}) derived for the low-temperature experimental crystal structure below 10 K and construct Hamiltonians on a 2D plane.  Here the parameters are listed in Table \ref{tab:abval} for the self-contained description. 
The interaction is screened by the interlayer screening, which must be taken into account when one solves a single layer Hamiltonian. It is established that this dimensional downfolding effect is well represented by a constant reduction of all the interactions by the amount $\Delta_{\rm DDF}$ from the values for the three dimensional Hamiltonian, if the subtracted value exceeds zero, and by truncation to zero if the subtracted value becomes negative~\cite{Nakamura2012}.
 For dmit compounds, it was estimated as $\Delta_{\rm DDF}=0.18 {\rm eV}$.
The interaction beyond the neighboring site (namely $V_4$ and beyond) becomes small after the dimensional downfolding and we ignore them. We then take into account up to the first neighbor transfer and interaction in each $\tilde{a},\tilde{b}$ and $\tilde{c}$ direction.  The exchange interaction is also ignored because it is not expected to play any visible role in the present study.
 In this work, we analyze the above Hamiltonian in the form of the lattice structure shown in Fig.~\ref{fig6} at half filling in the canonical ensemble, with $N=L\times L$ sites. 
We consider systems under the antiperiodic-periodic boundary condition.
 In all the cases of the compounds studied, the ground state is the Mott insulator.
\\

\noindent
{\bf Variational Monte Carlo method. }
In our simulations, we used a variational Monte Carlo method\cite{Tahara2008,MISAWA2019447,Nomura2017}. 
Our variational wave function takes the following form: 
\begin{eqnarray}
| \psi \rangle =  {\cal C} \ket{\phi}
\label{var_wf}
\end{eqnarray}
 with 
\begin{eqnarray}
{\cal C} = {\cal P}^{\rm G} {\cal P}^{\rm J_c} {\cal P}^{\rm J_s} {\cal M}^{\rm s} {\cal M}^{\rm c}. 
\label{C}
\end{eqnarray}
Here, ${\cal P}^{\rm G} = \exp \left( \sum_{i} \alpha_i^{\rm G} n_{i \uparrow} n_{i \downarrow} \right)$, ${\cal P}^{\rm J_c} = \exp \left( \sum_{i<j} \alpha_{ij}^{\rm J_c} n_{i} n_{j} \right)$ and 
${\cal P}^{\rm J_s} = \exp \left( \sum_{i,j} \alpha_{ij}^{\rm J_s} S^z_{i} S^z_{j} \right)$
are 
the Gutzwiller factor\cite{PhysRevLett.10.159}, the long-range Jastrow correlation factors\cite{PhysRev.98.1479, capello2005}, and the long-range spin Jastrow correlation factor\cite{PhysRevLett.60.2531}, respectively. 
Here, $S^z_i = n_{i\uparrow}-n_{i\downarrow}$.
In practice, we impose the translational symmetry on the Gutzwiller and Jastrow factors. 

To enhance the accuracy and to make the variance extrapolation explained in the next subsection easy, we also used two types of restricted Boltzmann machine (RBM) correlators\cite{Carleo2017,Nomura2017,Ferrari2019} ${\cal M}^{\rm s}$ and ${\cal M}^{\rm c}$, which are defined in the following equations:
\begin{equation}
{\cal M}^t = e^{a \sum_{i,\sigma} p^{t}_{i,\sigma}}\prod_{\alpha}^{N^t_\alpha} \prod_{r}^{N} 2 \cosh \left( b_\alpha + \sum_{i,\sigma} W_{i+r,\sigma,\alpha} p^{t}_{i,\sigma} \right),\label{rbm}
\end{equation}
\begin{equation}
p^{s}_{i,\sigma}= S_i^z =
    \begin{cases}
        1    &   \text{for $S_i^z\ket{x} = \ket{x}$}  \\
        0    &   \text{for $S_i^z\ket{x} = 0$}  \\
        -1   &   \text{for $S_i^z\ket{x} = -\ket{x}$}
    \end{cases},
\end{equation}
\begin{equation}
p^{c}_{i,\sigma}= 2(n_i-1)^2-1 =
    \begin{cases}
        1    &   \text{for $n_i\ket{x} \neq \ket{x}$}  \\
        -1   &   \text{for $S_i^z\ket{x} \neq 0$}
    \end{cases}.
\end{equation}
Here, $t$ in Eq.~(\ref{rbm}) represents a type of the RBM correlators, i.e. $t=c$ or $s$. $N^t_\alpha$ denotes the ratio of the number of the hidden units in the hidden layer to the number of the physical sites $N$.
$\ket{x}$ is a real space configuration of electrons in the sector where the total $S^z$ is zero. 
$a$, $\bm{b}$, and $\bm{W}$ are complex variational parameters.
For the measurements of the physical quantities defined in Methods~\ref{phys_quantities}, we use $N^c_\alpha =4$ and $N^c_\alpha = N^s_\alpha = 2$ for the nonmagnetic states and the antiferromagnetic state, respectively.

$| \phi \rangle$ in Eq.~(\ref{var_wf}) is a fermionic wavefunction. 
As $| \phi \rangle$, the generalized pairing wave function is employed, which is defined by 
\begin{eqnarray}
| \phi^{\rm pair} \rangle = \left( \sum_{i \sigma,j \sigma'} f_{i \sigma, j \sigma'} c_{i \sigma}^{\dagger} c_{j \sigma'}^{\dagger}  \right)^{N_{\rm e}/2} | 0 \rangle,\label{PPwf} 
\end{eqnarray}
where $f_{i \sigma, j \sigma'}$ are 
variational parameters and $N_{\rm e}$ is the total number of electrons. 
The spin indices $\sigma$ and $\sigma'$ can be arbitrary with singlet and triplet combinations.
This can 
accommodate the Hartree-Fock-Bogoliubov type wave function with antiferromagnetic (AF), charge and superconducting orders\cite{Tahara2008,giamarchi1991}, and flexibly describes these states and further paramagnetic metals as well. 
In this study, we extend it by introducing the dependence on the local density of $\ket{x}$. 
Our extended pairing wavefunction is defined as follows:
\begin{align}
& \ket{\phi^{\rm ext}} = \sum_x {\rm Pf} M(x) \ket{x}, \label{tcpair1}\\
& M_{nm}(x) = f^{\rm ext}_{\bm{r}_n \sigma, \bm{r}_m \sigma'}(x) - f^{\rm ext}_{\bm{r}_m \sigma', \bm{r}_n\sigma}(x),\label{tcpair2} 
\end{align}
\begin{equation}
    \label{tcpair3}
f^{\rm ext}_{i \sigma, j \sigma'}(x)=
    \begin{cases}
        f_{i \sigma, j \sigma'}^{\rm ss}    &   \text{for $n_{i,-\sigma}\ket{x} =0$, $n_{j,-\sigma'}\ket{x} =0$}  \\
        f_{i \sigma, j \sigma'}^{\rm sd}    &   \text{for $n_{i,-\sigma}\ket{x} =0$, $n_{j,-\sigma'}\ket{x} =\ket{x}$}  \\
        f_{i \sigma, j \sigma'}^{\rm ds}    &   \text{for $n_{i,-\sigma}\ket{x} =\ket{x}$, $n_{j,-\sigma'}\ket{x} =0$}  \\
        f_{i \sigma, j \sigma'}^{\rm dd}    &   \text{for $n_{i,-\sigma}\ket{x} =\ket{x}$, $n_{j,-\sigma'}\ket{x} =\ket{x}$}  
    \end{cases},
\end{equation}
where $n$ and $m$ are electron's indices in the sample $\ket{x}$. 
${\rm Pf} M$ is the Pfaffian of a skew-symmetric matrix $M$.
This extension improves accuracy of the fermionic part of the trial wavefunction especially for nonmagnetic states.
We treated $f_{i \sigma, j \sigma'}^{\rm ss}$, $f_{i \sigma, j \sigma'}^{\rm sd}$, $f_{i \sigma, j \sigma'}^{\rm ds}$, and $f_{i \sigma, j \sigma'}^{\rm dd}$ as complex variational parameters.

In order to reduce the computational cost by saving the number of independent variational parameters, we assume that $f^{\rm ext}_{ij}$ have a sublattice structure such that $f^{\rm ext}_{ij}$ depends on the relative vector ${\bm r}_i-{\bm r}_j$ and a sublattice index of the site $j$ which is denoted as $\eta(j)$. 
Thus, we can rewrite it as $f^{\rm ext}_{\eta(j)}({\bm r}_i-{\bm r}_j)$. 
In the present study on the 
nonmagnetic gapless states, we assumed a fully translational invariance (1$\times$1 sublattice structure) and do not optimize triplet pairings (represented by the case$\sigma=\sigma'$). 
For studies on the AF states, 
we extended the sublattice structure of $f^{\rm ext}_{ij}$ to $2 \times 2$. 
We did not use ${\cal P}^{\rm J_s}$ and ${\cal M}^{\rm s}$ for the optimization of the nonmagnetic states. 
All the variational parameters are simultaneously optimized by using the stochastic reconfiguration method\cite{sorella2001}.\\

\noindent
{\bf Variance extrapolation. }
\begin{figure}[h!]
  \begin{center}
    \includegraphics[width=8.5cm,clip]{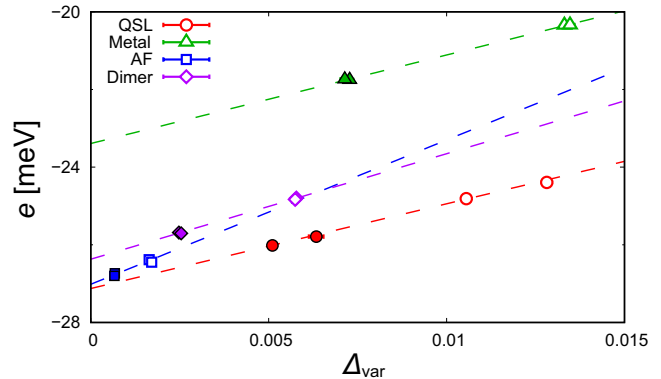}
  \end{center}
  \caption{
{\bf Example of the variance extrapolation of the energy to determine the ground state.} 
The example is shown for the $ab$ $initio$ Hamiltonian for $X$=EtMe$_3$Sb.
The variance is defined as $\Delta_{\rm var} = (\braket{H^2}-\braket{H}^2)/\braket{H}^2$.
$e$ represents the energy per site. 
Red circles, blue squares, green triangles and purple diamonds represent the energies of the QSL, the AF, a metal, and a spin-gap state with a finite dimer order along $x$ direction, respectively. Open and filled symbols are obtained by using the VMC method and the VMC combined with 1st-step Lanczos method, respectively, to each of which we also apply the restricted Boltzmann machine, which enables the lower energy. Dashed lines are fitted lines for the variance extrapolations. 
We impose the $1\times 1$ and $2 \times 1$ sublattice structures on the trial wavefunction for the metal and the dimer state, respectively.
In all the data plot, error bars obtained in the VMC calculation are smaller than the symbol size. }
\label{fig5}
\end{figure}
The AF and quantum spin liquid (QSL) states as well as the valence bond state are severely competing. 
To determine the lowest energy state among them, we performed extrapolations of energies to the limit of the zero energy variance~\cite{kwon1993,imada2000, sorella2001}. 
For this purpose, we combined with the restricted Boltzmann machine algorithm~\cite{Carleo2017,Nomura2017} together with the 1st Lanczos step\cite{heeb1993systematic}. 
To obtain the ground-state energy in the zero-variance limit, we perform the linear regression by using $N^c_\alpha =2$ and $4$ for nonmagnetic states. 
For the collinear antiferromagnetic state, we use $N^c_\alpha = N^s_\alpha = 0$ and $2$.
Examples of the extrapolations in the present studies are found in Fig.~\ref{fig5}, where
the variance dependence of the ground-state energy for the $ab$ $initio$ Hamiltonian for $X$=EtMe$_3$Sb is plotted.
The limit of the zero variance is our estimate of the ground state energy. 
We find severe competition between the AF and the QSL. The energies of the correlated metal and the dimer state with a finite spin gap are much higher than those of the AF and the QSL.
We note that the error bars in Fig. \ref{gspd} of the main text are obtained based on the propagation of errors as $\varepsilon = \sqrt{\varepsilon_{\rm AF}^2 + \varepsilon_{\rm QSL}^2}$,
where $\varepsilon_{\rm AF}$ and $\varepsilon_{\rm QSL}$ are the error arising from the linear regression for the AF and QSL states, respectively.\\

\noindent
{\bf Physical quantities. }\label{phys_quantities}
To elucidate the nature of the ground state, we analyze the following quantities: 
The primarily important quantity is the spin correlation 
\begin{eqnarray}
C({\bm r}) = \frac{1}{N_s} \sum_{\bm{r}'} \langle {\bm S}_{\bm{r}'} \cdot {\bm S}_{\bm{r}+\bm{r}'} \rangle 
\label{C(r)}
\end{eqnarray}
between the site $\bm r$ and $\bm{r}+\bm{r'}$
and its Fourier transform called the spin structure factor 
\begin{eqnarray}
S({\bm q}) = \frac{1}{N} \sum_{\bm{r},\bm{r}'} \langle {\bm S}_{\bm{r}} \cdot {\bm S}_{\bm{r}'} \rangle e^{i {\bm q}\cdot  ({\bm r}-{\bm r}')},
\label{S(q)}
\end{eqnarray}
where ${\bm S}_{\bm{r}}$  is the spin operator at the site $\bm{r}$. 

The spin Drude weight in $u$ direction ($u=x,y$) is defined by the second derivative of the energy with respect to the vector potential~\cite{Kohn1964}
\begin{eqnarray}
D^{u}_{\rm s} = \frac{1}{4\pi}\frac{\partial^2 E}{\partial A_u^2},
\label{Drude}
\end{eqnarray}
where $\bm{A}$ is inserted to the transfer by replacing as  
\begin{eqnarray}
t_{ij\sigma}\rightarrow t_{ij\sigma}\exp[i\sigma\pi \bm{A}\cdot (\bm{r}_i -\bm{r}_j)/L].
\label{VectPot}
\end{eqnarray}
The spin Drude weight is associated with the spin conductivity $\sigma(\omega)$ as
\begin{eqnarray}
D_{\rm s}=\int_{0}^{\Lambda}\sigma(\omega)
\label{Spin Conductivity}
\end{eqnarray}
with the cutoff $\Lambda$ to represent the weight around $\omega=0$.
If the peak around $\omega=0$ is given by the delta function, it is reduced to
\begin{eqnarray}
\sigma(\omega)=D_{\rm s}\delta(\omega).
\label{SpinConductivity}
\end{eqnarray}

We compute the statistical error of a physical quantity $O$ arising from the Monte Carlo sampling estimated as
\begin{eqnarray}
\sigma_{\rm err}(O) = \frac{1}{\sqrt{N_{\rm bin}}}\sqrt{ \frac{\sum_{i} [\braket{O}_i - \frac{1}{N_{\rm bin}} \left( \sum_{i} \braket{O}_i \right)]^2 }{N_{\rm bin}-1}}, \nonumber \\
\end{eqnarray}
where $\braket{O}_i$ is the expectation value of $O$ at the $i$-th bin in the Monte Carlo sampling.
In general, we use about $10^{6}$-$10^{7}$ Monte Carlo samples in each bin.
$N_{\rm bin}$ is the total number of bins and we typically set $N_{\rm bin}=5$-$10$.
\\

\noindent 
{\bf Gapless structure of wavefunction. }\label{app:gapless}
\begin{figure}[h!]
  \begin{center}
    \includegraphics[width=8cm,clip]{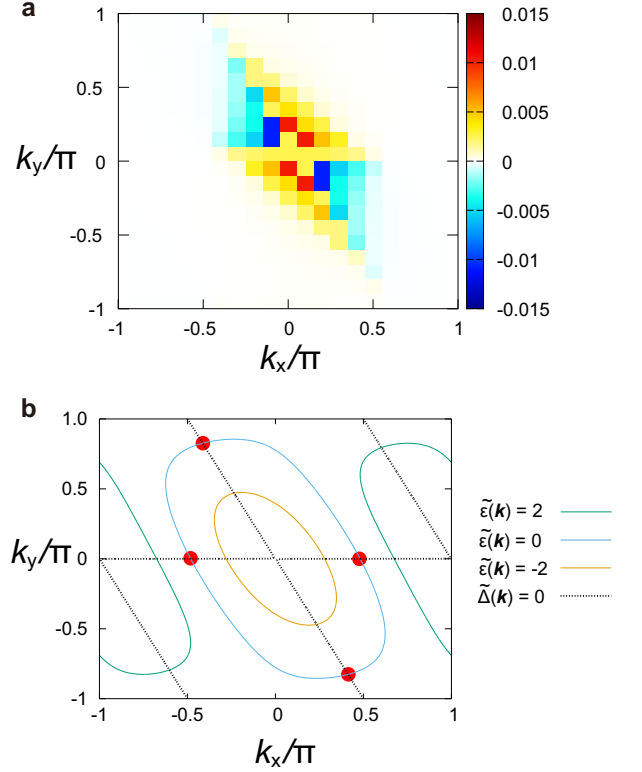}
  \end{center}
  \caption{
    {\bf Dispersion of excitation inferred from the optimized variational wave function.} The optimized $f(\bm {k})$ shown in {\bf a} are fitted by $\tilde{\epsilon}(\bm{k})$ and $\tilde{\Delta}(\bm{k})$ in the form Eq.~(\ref{BCSfij}) shown in {\bf b}. 
 The intersections of  the variational ``Fermi line" $\tilde{\epsilon}(\bm{k})=0$ (blue curve) and the vanishing gap line $\tilde{\Delta}(\bm{k})=0$ (dotted black lines), indicated by the red circles are the estimated gapless nodal points.}
\label{fitting}
\end{figure}

The method to analyze the structure of Fourier transform of the singlet pairing amplitude, $f_{ij}=f_{i\uparrow,j\downarrow}$, denoted by $f(\bm{k})$ is discussed here.
Although the correlation factors ${\cal C}$ in Eq.~(\ref{C}) 
and the density-dependent pairing amplitude in Eq. (\ref{tcpair1}) largely modify the wavefunction character, the nodal structure is expected to be governed by 
$f_{ij}$ in Eq.~(\ref{PPwf}) for $\sigma \neq \sigma'$.
The Fourier transform $f(\bm{k})$ can be interpreted as the solution of the BCS mean-field Hamiltonian 
\begin{eqnarray}
  {\cal H}_{\rm BCS}  =  -  \sum_{\bm{k},\sigma} \tilde{\epsilon}(\bm{k}) c_{\bm{k} \sigma}^{\dagger}c_{\bm{k} \sigma} +  \sum_k \tilde{\Delta}(\bm{k})  [c_{\bm{k} \sigma}^{\dagger} c_{-\bm{k} -\sigma}^{\dagger}+H.c] \nonumber \\
\label{BCSHamiltonian}
\end{eqnarray}
as~\cite{PhysRevLett.87.097201}
\begin{eqnarray}
  f(\bm{k})  =  \frac{\tilde{\Delta}(\bm{k})}{\tilde{\epsilon}(\bm{k})+\sqrt{\tilde{\epsilon}(\bm{k})^2+\tilde{\Delta}(\bm{k})^2}}.
\label{BCSfij}
\end{eqnarray} 
To obtain the fitted $f(\bm{k})$ defined in Eq.~(\ref{BCSfij}), we use the following form of $\tilde{\epsilon} (\bm{k})$ and $\tilde{\Delta} (\bm{k})$:
\begin{align}
\tilde{\epsilon} (\bm{k}) = -2\left( \cos k_x + \tilde{t_b} \cos (k_x + k_y) + \tilde{t_c} \cos k_y \right) - \tilde{\mu},\\
\tilde{\Delta} (\bm{k}) = 2\left( \tilde{\Delta_a}\cos k_x + \tilde{\Delta_b} \cos (k_x + k_y) + \tilde{\Delta_c} \cos k_y \right).
\end{align}
We also introduce the uniform scale factor $C$ of $f(\bm{k})$ because the original pairing amplitudes have the ambiguity for its norm.
The parameters $\tilde{t_b}, \tilde{t_c}, \tilde{\mu}, \tilde{\Delta_a}, \tilde{\Delta_b}$, $\tilde{\Delta_c}$ and $C$ are simultaneously optimized 
by using the differential evolution method\cite{Storn1997} implemented in SciPy\cite{2020SciPy-NMeth}, which is one of the global optimization methods.
We confirmed that the optimized results are similar even when we use other optimization methods such as the simplicial homology global optimization\cite{Endres2018}. 
We ignore the pairing amplitudes smaller than $10^{-5}$ during the optimization.
The fitting by $\tilde{\epsilon}(\bm{k})$ and $\tilde{\Delta}(\bm{k})$ reproduces the original data Fig.~\ref{fitting}(b) quite well and is not distinguishable in the color plot. 
Note that the original data are obtained from the optimized ${\cal P^{\rm G}}{\cal P^{\rm J_c}}\ket{\phi^{\rm pair}}$ for $\sigma \neq \sigma'$ with the real variational parameters $\alpha^{\rm G}_i,\alpha^{\rm J_c}_{ij}$ and $f_{ij}$. 

Figure~\ref{fitting}(b) shows the fitted $\tilde{\epsilon}(\bm{k})$ and $\tilde{\Delta}(\bm{k})$ by using Eq.~(\ref{BCSfij}). 
Since the excitation of the Hamiltonian is represented by $\sqrt{\tilde{\epsilon}(\bm{k})^2+\tilde{\Delta}(\bm{k})^2}$,
the gapless point appears at the cross points of $\tilde{\epsilon}(\bm{k})=0$ and $\tilde{\Delta}(\bm{k})=0$,
as are shown as red circles.
Though change in quantitative slopes of dispersion and broadening 
may take place, these nodal points 
may not be seriously altered by the correlation factors such as the Gutzwiller factor $\cal{P}^{\rm G}$\cite{Ferrari_2018}. 
The Fourier transform of $f_{ij}$ denoted by $f({\bm k})$ can be associated with the excitation spectra~\cite{PhysRevLett.87.097201} through the fitting to the ground state of a mean field BCS Hamiltonian with the $d$-wave type superconducting order.
Since the quasiparticle excitation of the BCS Hamiltonian corresponds to the spin-1/2 spinon, the excitation of the QSL is inferred to be characterized by the spin-$1/2$ Dirac-type spinon around the nodes of the $d$-wave superconducting state at around $(\pm\pi/2,0)$ and $(\pm\pi/2,\pi)$.
The spinon is an excitation resulted from the fractionalization of the spin and is confined. Measurable ordinary triplet excitations are given by a combination of two spinons generating the gapless points for the triplet excitations at around $(0,0)$, $(\pm\pi,0)$, $(0,\pm\pi)$, and $(\pm\pi,\pm\pi)$.\\

\noindent 
{\bf Strong coupling picture. }\label{app:strong_coupling}
\begin{table*}[]
\caption{
{\bf Effective exchange parameters evaluated from the {\it ab initio} parameters for $X$[Pd(dmit)$_2$]$_2$.} 
$J_n$ representes the effective coupling between the nearest neighbor sites along $n$ direction.
The unit of $J_n$ is meV. $J_b/J_a$, $J_c/J_a$, and $(J_c-J_b)/J_a$ are also listed.
}
\begin{center}
\scalebox{1.25}{
\footnotesize

\begin{tabular}{l|ccc|ccc}
\hline \hline
Cation $X$   &  $J_a$ & $J_b$ & $J_c$ & $J_b/J_a$& $J_c/J_a$ & $(J_c-J_b)/J_a$ \\
\hline 
 Me$_4$P & 38.8 & 29.3  & 9.17 & 0.76 & 0.24 & -0.52  \\ 
 Me$_4$As & 41.3 & 26.5 & 9.04 & 0.64 & 0.22 & -0.42\\ 
 Me$_4$Sb & 30.5 & 21.1 & 11.7 & 0.69 & 0.38 & -0.31\\
 Et$_2$Me$_2$As & 29.4 & 19.1 & 12.2 & 0.65 & 0.41 & -0.23\\
 EtMe$_3$Sb & 30.5 & 19.6 & 14.4 & 0.64 & 0.47 & -0.17 \\
\hline \hline
\end{tabular}
}
\end{center}
\label{tab:exchange}

\end{table*}
The effective exchange coupling in the strong coupling expansion in terms of either $t_a/(U-V_1)$, $t_b/(U-V_2)$ or $t_c/(U-V_3)$ can be easily derived using $J_a=4t_a^2/(U-V_1)$ etc., which is summarized in Table \ref{tab:exchange}. For $X$=EtMe$_3$Sb, 
 $J_a=30.5$ meV, $J_b= 19.6$ meV and $J_c=14.4$ meV (we ignore the contribution from the direct exchange because its small and more or less isotropic values less than 3meV~\cite{Misawa2021}, which  have only minor effect): The one-dimensionality is in fact as a consequence of not $J_b/J_a$ (or $J_c/J_a$) but small $(J_c-J_b)/J_a$. Nonetheless, this factor is 0.17, which is substantially larger than the value by Kenny {\it et al.}, 0.053~\cite{PhysRevMaterials.4.044403}. 
In fact, with our present {\it ab initio} parameters, their way of estimate of the phase diagram using the above $J_a, J_b, J_c$ and their estimate of the ring exchange interaction $K_4=80t_b^2t_c^2/(U-(V_1+V_2+V_3)/3)^3$ giving $K_4/J_a\sim 0.11$ indicate that the EtMe$_3$Sb[Pd(dmit)$_2$]$_2$ is located near the border of the antiferromagnetic phase but in the QSL side, consistently with our full quantum calculation. 
The estimates of the ground states for the other compounds based on their criterion are also consistent with our results. However, their recent estimate of the instability to the antiferromagnetic order contradicts our results\cite{PhysRevMaterials.5.084412}. It could be related to the neglect of the ring exchange interaction and partly because of the limitation of the random phase approximation based on the quasi-one-dimensionality employed by them~\cite{PhysRevMaterials.5.084412}. 
In any case, it is clear that the present analysis is quantitatively the most comprehensive and accurate analysis because all orders of expansion even beyond the ring exchange from the viewpoint of the strong coupling expansion are included and the present original itinerant-electron treatment is the most strict first principles analysis.\\

\noindent
{\bf Acknowledgments}

The authors thank Reizo Kato for discussions and clarifications on the experimental results
{and Shigeki Fujiyama for notifying his unpublished result on the coexistence of AF and QSL phases for $\beta'$-Et$_2$Me$_2$As[Pd(dmit)$_2$]$_2$ and $\beta'$-Me$_4$Sb[Pd(dmit)$_2$]$_2$.}
KI thanks Yusuke Nomura for useful comments and the implementation on the restricted Boltzmann machine correlator.
TM and KY thank Takao Tsumuraya for discussions on the derivation of
the $ab$ $initio$ Hamiltonians. This work was supported in part by KAKENHI Grant No. 16H06345 and 19K14645 from JSPS. 
This research was also supported by MEXT as ``program for Promoting Researches on the Supercomputer Fugaku"(Basic Science for Emergence and Functionality in Quantum Matter - Innovative Strongly Correlated Electron Science by Integration of Fugaku and Frontier Experiments -, JPMXP1020200104). 
We thank the Supercomputer Center, the Institute for Solid State Physics, The University of Tokyo for the use of the facilities.
We also thank the computational resources of supercomputer Fugaku provided by the RIKEN Center for Computational Science (Project ID: hp200132 and hp210163) and Oakbridge-CX in the Information Technology Center, The University of Tokyo. \\


\end{document}